\documentclass[conference,a4paper]{IEEEtran}
\IEEEoverridecommandlockouts
\ifCLASSINFOpdf
 \usepackage[pdftex]{graphicx}
\else
 \usepackage[dvips]{graphicx}
\fi

\usepackage[cmex10]{amsmath}
\usepackage{cite}
\usepackage{algorithmic}
\usepackage{algorithm}
\usepackage{mdwmath}
\usepackage{mdwtab}
\usepackage{amssymb}
\usepackage{array}
\usepackage{multirow}
\usepackage{mdwlist}
\usepackage[usenames,dvipsnames]{color}

\begin{document}

\title{Uplink Non-Orthogonal Multiple Access for 5G Wireless Networks}

\author{\IEEEauthorblockN{Mohammed~Al-Imari, Pei~Xiao, Muhammad~Ali~Imran, and Rahim~Tafazolli}
\IEEEauthorblockA{Centre for Communication Systems Research \\University of Surrey\\
Guildford, GU2 7XH, United Kingdom\\
Email:\{m.al-imari, p.xiao, m.imran, r.tafazolli\}@surrey.ac.uk}}

\maketitle
\begin{abstract}
Orthogonal Frequency Division Multiple Access (OFDMA) as well as other orthogonal multiple access techniques fail to achieve the system capacity limit in the uplink due to the exclusivity in resource allocation. This issue is more prominent when fairness among the users is considered in the system. Current Non-Orthogonal Multiple Access (NOMA) techniques introduce redundancy by coding/spreading to facilitate the users' signals separation at the receiver, which degrade the system spectral efficiency. Hence, in order to achieve higher capacity, more efficient NOMA schemes need to be developed. In this paper, we propose a NOMA scheme for uplink that removes the resource allocation exclusivity and allows more than one user to share the same subcarrier without any coding/spreading redundancy. Joint processing is implemented at the receiver to detect the users' signals. However, to control the receiver complexity, an upper limit on the number of users per subcarrier needs to be imposed. In addition, a novel subcarrier and power allocation algorithm is proposed for the new NOMA scheme that maximizes the users' sum-rate. The link-level performance evaluation has shown that the proposed scheme achieves bit error rate close to the single-user case. Numerical results show that the proposed NOMA scheme can significantly improve the system performance in terms of spectral efficiency and fairness comparing to OFDMA.
\end{abstract}

\IEEEpeerreviewmaketitle
\section{Introduction}
The explosive traffic growth in mobile communications has motivated research activities, in both academic and industrial communities, to design the next generation (5G) of wireless networks that can offer significant improvements in coverage and user experience~\cite{5GEditorial14}. 5G wireless networks demand highly spectral-efficient multiple access techniques, which play an important role in determining the performance of mobile communication systems. In general, multiple access techniques can be classified into orthogonal and non-orthogonal based on the way the resources are allocated to the users~\cite{OMAvsNOMA}. In Orthogonal Multiple Access (OMA) techniques, the users in each cell are allocated the resources exclusively and there is no inter-user interference, hence, low-complexity detection approaches can be implemented at the receiver to retrieve the users' signals. In the current mobile communication systems (such as Long-Term Evolution (LTE) and LTE-Advanced~\cite{LTEA_Ghosh}), OMA techniques have been adopted, e.g. Orthogonal Frequency Division Multiple Access (OFDMA) and Single Carrier Frequency Division Multiple Access (SC-FDMA). On the other hand, in Non-Orthogonal Multiple Access (NOMA) techniques, all the users can use resources simultaneously, which leads to inter-user interference. Consequently, more complicated Multi-User Detection (MUD) techniques are required to retrieve the users' signals at the receiver.\\
\indent Theoretically, orthogonal transmission is suitable for downlink as it can achieve the maximum users' sum-rate~\cite{Trans_Por_Adap}. Also, it is more difficult to implement MUD techniques at the user equipment due to the limited processing power. In uplink, OMA is not optimal in terms of spectral efficiency, and cannot achieve the system upper bound~\cite{OFDMA_Optimality}. This issue is more prominent when fairness among the users is considered in the system~\cite{Tse_Polymatroid}. Thus, to improve the system spectral efficiency, NOMA techniques need to be adopted in next generation of wireless networks. The optimal approach for NOMA is to allow all the users to share each resource element (in frequency or time domain), and the users' power allocated through iterative water-filling~\cite{WeiYu_IWFJour}. However, in the optimal NOMA scheme, there is no control on the number of users that share each subcarrier, which makes the MUD at the receiver infeasible. Other techniques that allow NOMA (such as Code Division Multiple Access (CDMA), Interleave Division Multiple Access (IDMA)~\cite{IDMA_LiPing}, Low Density Spreading (LDS)~\cite{LDS_Editorial}, etc.) add redundancy via coding/spreading to facilitate the users separation at the receiver. However, the introduced redundancy will inevitably degrade the system spectral efficiency~\cite{CodSpread_tradeoff}.\\
\indent In this paper, we propose an uplink NOMA scheme for Orthogonal Frequency-Division Multiplexing (OFDM) without coding/spreading redundancy. The users will use the subcarriers without any exclusivity, and at the receiver MUD will be implemented for users' separation. However, to control the receiver complexity, the number of users in each subcarrier will be limited to a specific number. The main advantages of the proposed NOMA scheme are: 1) Higher spectral efficiency comparing to current OMA and NOMA techniques. 2) Lower receiver complexity comparing to optimal-unconstrained NOMA scheme. In addition, subcarrier and power allocation algorithm for the new NOMA scheme will be proposed. An evaluation of the link-level and system-level performance of the proposed scheme will be carried out.\\
\indent This paper is organized as follows: Sec.~\ref{SectionTwo} presents the proposed NOMA scheme along with the link-level performance evaluation. In Sec.~\ref{SectionThree}, we propose subcarrier and power allocation algorithm for the new NOMA scheme. In Sec.~\ref{SectionFour}, we evaluate the system-level performance of the proposed scheme in terms of spectral efficiency and fairness. Finally, concluding remarks are drawn in Sec.~\ref{SectionFive}.
\section{Background and System Model}\label{SectionTwo}
\begin{figure}[!t]
\centering
\includegraphics[width=8.87cm]{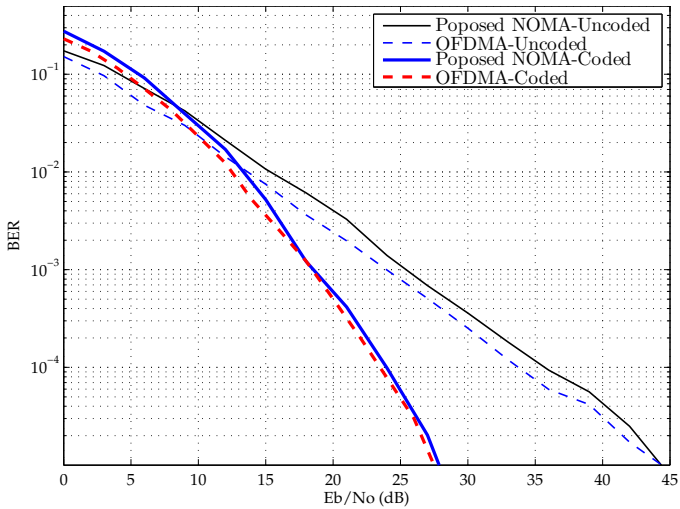}
\caption{BER comparison for the proposed NOMA (with $L=2$) scheme and OFDMA.}
\label{fig_BER_Results}
\end{figure}

The uplink single-cell scenario can be represented by a generic Multiple Access Channel (MAC), where a set of users $\mathcal{K}=\{1,\cdots, K\}$ transmit to a single base station in the presence of additive Gaussian noise over frequency selective channels~\cite{Tse_Polymatroid}. The available bandwidth is divided into a set of subcarriers $\mathcal{N}=\{1,\cdots, N\}$, and a user $k\in\mathcal{K}$ can transmit over the subcarriers, with transmission power $p_{k,n}$ over subcarrier $n\in\mathcal{N}$ subject to individual maximum power constraints $P_k:\sum_{n\in\mathcal{N}} p_{k,n} \leq P_k$. Using Shannon capacity formula, the $k$th user rate $(R_k)$ is given by
\begin{equation}\label{Eq_UserRate}
R_k = \sum_{n\in\mathcal{N}} \log_2{\left(1+\frac{p_{k,n}h_{k,n}}{\sigma^2+I_{k,n}}\right)},\quad\text{bit/s/Hz},
\end{equation}
where $h_{k,n}$ is the channel gain of the $k$th user on the $n$th subcarrier, $\sigma^2$ in the noise power per subcarrier, and $I_{k,n}$ is the interference the $k$th user will see on subcarrier $n$ from other users. The users' power allocation over the subcarriers can be optimized to maximize certain system objective, such as spectral efficiency and fairness. In the generic MAC, users' total transmit power $(P_k)$ is the only constraint on the power allocation optimization problem, and there is no constraint on the number of users that share each subcarrier. In this case, the optimal power allocation that maximizes the users' sum-rate can be achieved through iterative water-filling~\cite{WeiYu_IWFJour}. Clearly, without any constraint on the number of users per subcarrier, a large number of users may be active on some subcarriers. Consequently, the receiver complexity to separate the users' signals will be very high, rendering the generic MAC infeasible for practical systems.\\
\indent For practical multiple access schemes, additional constraints are imposed to control the number of users that share the same subcarrier. Let $\mathcal{S}_n$ be the set of active users at the $n$th subcarrier, i.e., $\mathcal{S}_n=\{k: p_{k,n}>0\}$. As a special multiple access scheme, in OMA (e.g. OFDMA and SC-FDMA), no more than one user can use the same subcarrier at the same time. In other words, any OMA scheme can be defined mathematically as $\vert \mathcal{S}_n\vert\in\{0, 1\},\quad\forall n\in\mathcal{N}$. This constraint will ensure orthogonality among the users and eliminates the inter-user interference, hence, single-user detection can be used to retrieve the users' signals. However, this exclusivity in subcarrier allocation reduces the system's spectral efficiency and fairness, as the subcarriers allocated to one user cannot be used by other users. In order to maximize the sum-rate without the fairness consideration, the users close the base-station will be allocated most of the subcarriers and the cell-edge users will not be able transmit. More detailed analysis of the OMA sub-optimality can be found in~\cite{OFDMA_Optimality}.\\
\indent On the other hand, for NOMA schemes, this constraint is more relaxed as more than one user can share the same subcarrier, i.e., $\vert \mathcal{S}_n\vert$ can be greater than 1. However, allowing the users share the same resources requires MUD at the receiver. In current NOMA schemes (CDMA, IDMA, LDS, etc.), redundancy is added to the users signals through spreading or coding to allow low complexity MUD techniques. Clearly, the spreading/coding used to facilitate the users separation reduces the system spectral efficiency~\cite{CodSpread_tradeoff}.\\
\indent Under optimal power allocation in generic MAC, each user will use only the subcarriers with good channel conditions, and the subcarriers with bad channel conditions will have zero power. Hence, as the users experience different and independent fades, it is very unlikely that all the users will use all the subcarriers. However, some subcarriers may be overloaded with many users active on them, which make the receiver complexity to be infeasible. Inspired by these facts, here we propose a new NOMA technique, in which the users will be able to share the subcarriers, however, an upper limit on the number of users per subcarrier will be enforced. Let $L$, be the maximum number of users that are allowed to share each subcarrier, the proposed scheme can be defined as
\begin{equation}
\vert \mathcal{S}_n\vert\leq L,\quad\forall n\in\mathcal{N}.
\end{equation}
By setting $L\ll K$, the receiver complexity can be significantly reduced. Hence, at the receiver, optimum MUD techniques can be implemented with moderate complexity, without the need for spreading/coding redundancy that reduce the system spectral efficiency. The complexity of the optimum MUD for the proposed scheme will be $\mathcal{O}(|\mathcal{X}|^L)$, where $\mathcal{X}$ denotes the constellation alphabet, which is significantly reduced compared to a complexity order up to $\mathcal{O}(|\mathcal{X}|^K)$ for optimum MUD with generic MAC. The number of users per subcarrier can be changed based on the system loading to balance between the system spectral efficiency and the receiver complexity.\\
\begin{figure}[!t]
\centering
\includegraphics[width=8.87cm]{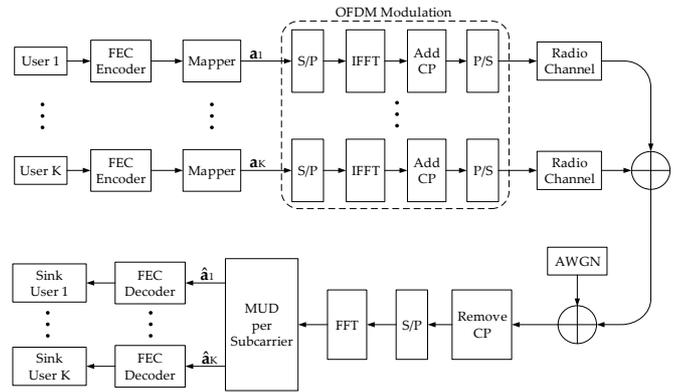}
\caption{Block digram of the proposed NOMA technique.}
\label{fig_PNOMA}
\end{figure}
\indent Comparing to OMA, the proposed scheme achieves higher spectral efficiency by allowing more than one user to share the same subcarrier. Comparing to the existing NOMA schemes, no redundancy (spreading/coding) is used in the proposed scheme, hence, more efficient utilization of the resources can be achieved. Furthermore, the proposed scheme is more practical to implement in real system in comparison to the generic MAC. The conceptual block diagram of the proposed uplink NOMA scheme is depicted in Fig.~\ref{fig_PNOMA}. Let $\textbf{a}_k\in\mathcal{X}^{M_k}$ be the symbols vector of user $k$ consisting of $M_k$ modulated symbols, where $a_{k,n}\in\mathcal{X}$ is the transmitted symbol of the $k$th user on the $n$th subcarrier. As no spreading is implemented in the proposed NOMA scheme, the transmitter side will be the same as the one for OFDMA. However, as the users are not restricted to exclusively use the subcarriers, at the receiver, the symbols from $L$ different users will be superimposed. Hence, after performing OFDM demodulation operation at the receiver, the received signal on the $n$th subcarrier will be
\begin{equation}
y_n=\sum_{k\in\mathcal{S}_n} a_{k,n}\sqrt{h_{k,n}}+z_n,
\end{equation}
where $z_n\sim \mathcal{N}(0,\sigma^2)$ is the Gaussian random noise on the $n$th subcarrier. The received signal $y_n$ is passed to MUD for signal detection. The optimal MUD, that achieves the optimal performance in terms of error probability, uses the Maximum Likelihood (ML) criterion in detecting the users symbols, which selects the symbols sequence, $\mathbf{\hat{a}}_n\in\mathcal{X}^L$, that maximizes the likelihood function $p(y_n|\mathbf{a}_n)$ given the channel observation
\begin{equation}\label{Eq_ML_MUD}
\mathbf{\hat{a}}_n=\arg\max_{\mathbf{a}_n\in\mathcal{X}^L}\bigg(-\lVert y_n-\mathbf{a}_n\mathbf{h}_n^T\rVert^2\bigg).
\end{equation}
Here, $\mathbf{a}_n$ and $\mathbf{h}_n$ are the vectors that contain the symbols transmitted on the $n$th subcarrier and their corresponding channel gains, respectively. Then, each user estimated symbols will be passed to the channel decoder. The link-level performance of the proposed scheme is evaluated and compared with OFDMA in Fig.~\ref{fig_BER_Results}, where we show the Bit Error Rate (BER) versus $Eb/N_0$ (energy per bit to noise power spectral density ratio) for the proposed NOMA and OFDMA systems with BPSK modulation. For NOMA, two users per subcarrier is used (i.e., $L=2$) and ML-MUD is implemented. Total number of subcarriers of $64$ with $15$KHz subcarrier spacing is assumed. ITU pedestrian B channel model~\cite{IMT_Guideline} is adopted for generating fast fading, and a half-rate convolutional code is used for the coded case. As the figure shows, for the uncoded case the proposed NOMA scheme experience small BER degradation comparing to OFDMA. However, this performance loss is almost vanish when coding is used. The results suggest that for small number of users per subcarrier, the link-level of the proposed NOMA scheme is close to the single-user case (no inter-user interference).
\section{Subcarrier and Power Allocation}\label{SectionThree}
In this section, the subcarrier and power allocation for the proposed NOMA scheme will be considered. Firstly, the subcarrier and power allocation problem is formulated as sum-rate maximization. Then, a suboptimal subcarrier and power allocation algorithm will be proposed.
\subsection{Problem Formulation}
Here, we assume that the base station obtain the Channel State Information (CSI), and based on the obtained CSI, the base station assigns subcarriers and power to each user to optimize the system performance. To formulate the subcarrier and power allocation problem, let $x_{k,n}$ be the channel allocation index such that $x_{k,n}=1$ if subcarrier $n$ is allocated to user $k$ and $x_{k,n}=0$ otherwise. As the users' signals superimposed at the receiver, successive decoding will be assumed to perform the joint processing. It is worth mentioning that the decoding order does not affect the sum-rate, and any arbitrary decoding order can be assumed. Accordingly, it will assumed that the users are decoded in an increasing order of their indices. Hence, the first user to be decoded,~$k=1$, will see interference from all the other users $k=2, \cdots, K$, and the second user to be decoded will see interference from the users $k=3, \cdots, K$, and so on. Thus, the interference $(I_{k,n})$ each user experience on each subcarrier with this decoding order will be
\begin{equation}\label{Eq_ScU_Interf}
I_{k,n} = \sum_{j=k+1}^K x_{j,n} p_{j,n} h_{j,n},\quad k=1, \cdots, K-1.
\end{equation}
Consequently, the subcarrier and power allocation can be formulated as follows
\begin{equation}\label{Eq_SPAobj}
\max_{x_{k,n},p_{k,n}} \sum_{k\in\mathcal{K}} \sum_{n\in\mathcal{N}}x_{k,n} \log{\left(1+\frac{p_{k,n}h_{k,n}}{I_{k,n}+\sigma^2}\right)},
\end{equation}
subject to
\begin{equation}\label{Eq_LoadingCons}
\sum_{k\in\mathcal{K}} x_{k,n} \leq L,\quad \forall n\in\mathcal{N},
\end{equation}
\begin{equation}\label{Eq_BinCons}
x_{k,n}\in \{0,1\}, \quad\forall k\in\mathcal{K},\hspace{0.2cm} n\in\mathcal{N},
\end{equation}
\begin{equation}\label{Eq_MaxPwrCons}
\sum_{n\in\mathcal{N}} p_{k,n} \leq P_k,\quad \forall k\in\mathcal{K},
\end{equation}
\begin{equation}\label{Eq_PostPwrCons}
p_{k,n}\geq 0, \quad \forall k\in\mathcal{K},\hspace{0.2cm} n\in\mathcal{N}.
\end{equation}
The constraint in~\eqref{Eq_LoadingCons} provides control on the number of users per subcarrier. Unfortunately, this optimization problem cannot be expressed as a convex one for two reasons. Firstly, the binary constraint in~\eqref{Eq_BinCons} is a non-convex set, and secondly, the interference term in the objective function~\eqref{Eq_SPAobj} makes it a non-convex function. Unlike the case in OFDMA, the constraint in \eqref{Eq_BinCons} cannot be relaxed to take any real value in the interval $[0, 1]$ to make the problem tractable~\cite{EURASIP_OFDMA}. In OFDMA, when the constraint is relaxed the users are still orthogonal to each other. If the binary constraint is relaxed for the proposed NOMA scheme, all the users may interfere on each subcarrier, which violate the main concept of our proposed scheme that only $L$ users are interfering on each subcarrier.
\subsection{Suboptimal Algorithm}
As it has been discussed, finding the optimal solution to the subcarrier and power allocation problem (\ref{Eq_SPAobj}~-~\ref{Eq_PostPwrCons}) is intractable and impractical. Therefore, suboptimal subcarrier and power allocation algorithm with low complexity is presented here. Let $\mathcal{N}_k^u$ represents the set of subcarriers that unallocated to the $k$th user and not allocated to more than $L$ other users, and let $\mathcal{N}_k^a$ be the set of subcarriers allocated to that user. The proposed algorithm consists of three steps: 1) \emph{Power Allocation}: each user performs Single-User Water-Filling (SUWF) over all the available subcarriers ($\mathcal{N}_k^a\cup\ \mathcal{N}_k^u $), considering the interference from other users. Based on the power allocation, the users' rates over the available subcarriers are calculated. 2) \emph{Subcarrier Selection}: for each user, find the subcarrier that has the maximum rate \emph{(selected subcarrier)} among the unallocated subcarriers $(\mathcal{N}_k^u)$. 3) \emph{Subcarrier Allocation}: based on the users' rates, one subcarrier (of the \emph{selected subcarriers}) is allocated to one user. Two subcarrier allocation approaches will be considered. The first one (which is referred to as Local Rate Maximization (LRM)) is allocating a subcarrier to the user that has the maximum rate on its \emph{selected subcarrier}. The second approach (which is referred to as Global Objective Maximization (GOM)) is to allocate a subcarrier to the user that achieve the maximum increase in the objective function~\eqref{Eq_SPAobj}
\begin{equation}
k^\star=\arg\max_k (R_k-R_k^a),
\end{equation}
where $R_k^a$ is the rate of user $k$ using the subcarriers that already allocated to that user, $\mathcal{N}_k^a$. The algorithm will iteratively allocate the subcarriers one by one until all the subcarriers reaches there maximum limit $(L)$ or the users' rate cannot be increased more. The detailed algorithm is listed in Algorithm~\ref{alg_RRM}.
\begin{algorithm}[!t]
\small
\caption{\small Iterative Subcarrier and Power Allocation}
\label{alg_RRM}
\begin{algorithmic}[1]
\STATE \textbf{Initialization:} Put $I_{k,n}=0$, $\mathcal{N}_k^a=\emptyset$ and $\mathcal{N}_k^u=\mathcal{N},\hspace{0.3cm}\forall k\in\mathcal{K}$.
\REPEAT
\STATE \textbf{Power Allocation:} Considering the interference from other users, each user performs SUWF over $(\mathcal{N}_k^a\cup \mathcal{N}_k^u)$.
\STATE \textbf{Subcarrier Selection:} Find the best subcarrier $(b_k)$ for each user: $b_k=\arg\displaystyle\max_{n\in\mathcal{N}_k^u} R_{k,n},\quad\forall k\in\mathcal{K}$.
\STATE \textbf{Subcarrier Allocation:} \\
\textbf{LRM:} $k^\star=\arg\displaystyle\max_{k\in\mathcal{K}} R_{k,b_k}$.\\
\textbf{GOM:} $k^\star=\arg\displaystyle\max_{k\in\mathcal{K}} (R_k-R_k^a)$
\STATE Allocate the subcarrier $b_{k^\star}$ to user $k^\star$:\\
Set $x_{k^\star,b_{k^\star}}=1,\quad\mathcal{N}_{k^\star}^a=\mathcal{N}_{k^\star}^a\cup \{n\}$.
\IF{$\sum_{k\in\mathcal{K}} x_{k,b_{k^\star}} = L,$}
\STATE $\mathcal{N}_k^u=\mathcal{N}_k^u\setminus{b_{k^\star}},\quad\forall k\in\mathcal{K}$.
\ENDIF
\STATE Update the interference $(I_{k,n})$ based on the current subcarrier allocation.
\UNTIL$R_{k,n}=0,\ \forall n\in\mathcal{N}_k^u, k\in\mathcal{K}$ or $\mathcal{N}_k^u=\emptyset,$ $\quad\forall k\in\mathcal{K}$.
\end{algorithmic}
\end{algorithm}
\vspace{-0.3cm}
\section{Numerical Results}\label{SectionFour}
\begin{figure}[!t]
\centering
\includegraphics[width=8.87cm]{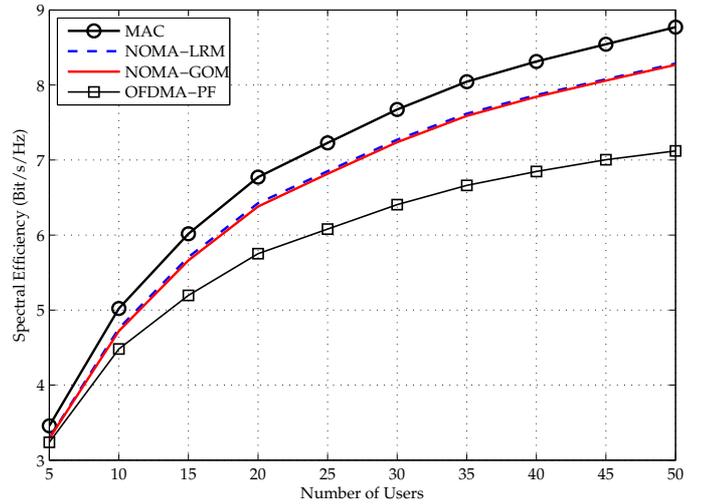}
\caption{Spectral efficiency evaluation of the proposed NOMA technique.}
\label{fig_SE_Results}
\end{figure}

\begin{figure}[!t]
\centering
\includegraphics[width=8.8cm]{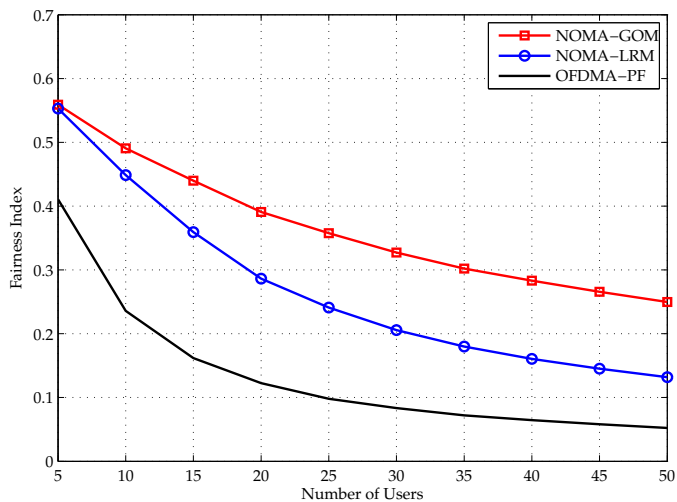}
\caption{Fairness comparison between the proposed NOMA technique and OFDMA.}
\label{fig_FI_Results}
\end{figure}

\begin{figure}[!t]
\centering
\includegraphics[width=8.8cm]{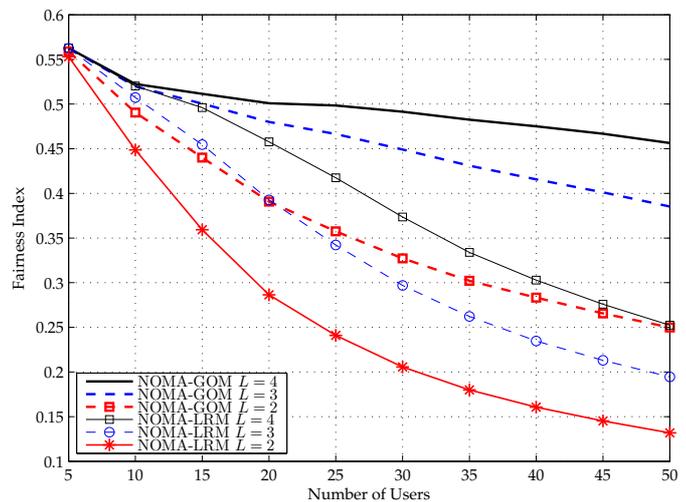}
\caption{Fairness comparison for different subcarrier loading $(L)$.}
\label{fig_FI_Comp}
\end{figure}

In this section, the system-level performance of the proposed NOMA with subcarrier and power allocation is evaluated through Monte Carlo simulation. A single-cell with $0.5$~km radius is considered, where the users' locations are randomly generated and uniformly distributed within the cell. The maximum transmit power of each user is $23$~dBm and the system bandwidth is $10$~MHz consisting of $50$~resource blocks. ITU pedestrian B fast fading model and the COST231 Hata propagation model for microcell environment~\cite{TR25_996} are adopted. Lognormal shadowing with $8$~dB standard deviation is implemented. The noise power spectral density is assumed to be $-173$~dBm/Hz. The generic MAC with iterative water-filling~\cite{WeiYu_IWFJour} and OFDMA with proportional fair (PF) subcarrier and power allocation are used for benchmark comparison. The generic MAC with iterative water-filling represents the upper bound on the system's spectral efficiency. Fig.~\ref{fig_SE_Results} shows the spectral efficiency comparison versus the total number of users in the system $(K)$, with maximum two users per resource block for the proposed NOMA scheme. As it can be seen from the figure, the proposed NOMA significantly outperforms OFDMA (as OMA technique) and achieves spectral efficiency that is closer to the system upper bound (generic MAC). As the total number of users is increased, the gap between the performance of OFDMA and the system upper bound becomes larger. On the contrary, NOMA is able to keep an acceptable performance level. For example, at $K=50$, the proposed NOMA scheme achieves about $95\%$ of the system upper bound, while OFDMA only achieves $81\%$. Also, it can be noticed that the two subcarrier allocation criteria (LRM and GOM) achieve almost the same spectral efficiency performance.
To evaluate the fairness of the proposed NOMA scheme, Fig.~\ref{fig_FI_Results} shows the Jain's fairness index for the proposed NOMA and OFDMA systems. The Jain's fairness index is given by~\cite{Jain_Fairness}\vspace{-0.2cm}
\begin{equation}
\text{Jain's fairness index}= \frac{(\sum_{k=1}^K R_k)^2}{K\sum_{k=1}^K R_k^2}.
\end{equation}
Jain's fairness index is bounded between 0 and 1 with the maximum achieved by equal users' rates. As the figure shows, the proposed NOMA scheme (with both subcarrier allocation criteria) is fairer comparing to OFDMA, even though proportional fair subcarrier and power allocation is implemented for OFDMA. This is due to the fact that in OFDMA when a resource block is allocated to one user, it cannot be allocated to other users, thus, some users will end up having no resource blocks allocated or with very small rate. This performance gap between the two systems is more significant at high number of users. On the other hand, in the proposed NOMA scheme, there is no exclusivity constraint, and the resource blocks can be reused by other users within the same cell. Hence, more users can be supported in the system. Furthermore, it can be observed that the subcarrier allocation criterion GOM is considerably fairer comparing to the LRM criterion. Thus, as both subcarrier allocation criteria achieve the same spectral efficiency, GOM can be considered superior to LRM. However, GOM approach is relatively more computationally complex comparing to LRM.\\
\indent In order to quantify the effect of the subcarrier loading $(L)$, Fig.~\ref{fig_FI_Comp} shows the Jain's fairness index for the proposed NOMA technique with $L=2, 3, \text{and}\ 4$. As the results show, by allowing more users to share the subcarriers, better fairness levels can be achieved. However, this will increase the receiver complexity, thus, selecting the maximum number of users per subcarrier need to be chosen carefully to keep the complexity affordable. On the other hand, increasing the number of users per subcarrier results in marginal increase in spectral efficiency. Due to space limitation, the spectral efficiency comparison is not presented here. Considering the results altogether, it can concluded that the proposed NOMA technique achieves better system performance in terms of spectral efficiency and fairness comparing to OFDMA. Also, the spectral efficiency achieved by the proposed NOMA is relatively closer to the system upper bound.
\vspace{-0.4cm}
\section{Conclusion}\label{SectionFive}
5G wireless networks demand highly spectral-efficient multiple access techniques to meet the ever-increasing traffic growth in mobile communications. In this paper, we proposed a novel uplink Non-Orthogonal Multiple Access (NOMA) technique, in which the users are able to share the subcarriers. Optimum multi-user detection is implemented at the receiver to separate the users' signals. An upper limit on the number of users per subcarrier have been imposed to control the receiver complexity. Despite the inter-user interference, the proposed NOMA scheme have shown to achieve link-level performance that is very close to the single-user case. In addition, we have developed subcarrier and power allocation algorithm for the proposed NOMA scheme. The system-level performance of the proposed NOMA scheme is investigated and simulation results have shown that the proposed scheme significantly improves the system spectral efficiency and fairness comparing to orthogonal multiple access. Thus, the proposed NOMA scheme can be considered as promising candidate for next generation wireless networks. Future work will include further study and investigation to evaluate the link-level performance of the proposed NOMA scheme with different multi-user detection and decoding techniques.
\section*{Acknowledgment}
We would like to acknowledge the support of the University of Surrey 5GIC members for this work.
\bibliographystyle{IEEEtran}
\bibliography{IEEEfull,References_NOMA_ISWCS14}

\begin{thebibliography}{10}
\providecommand{\url}[1]{#1}
\csname url@samestyle\endcsname
\providecommand{\newblock}{\relax}
\providecommand{\bibinfo}[2]{#2}
\providecommand{\BIBentrySTDinterwordspacing}{\spaceskip=0pt\relax}
\providecommand{\BIBentryALTinterwordstretchfactor}{4}
\providecommand{\BIBentryALTinterwordspacing}{\spaceskip=\fontdimen2\font plus
\BIBentryALTinterwordstretchfactor\fontdimen3\font minus
  \fontdimen4\font\relax}
\providecommand{\BIBforeignlanguage}[2]{{%
\expandafter\ifx\csname l@#1\endcsname\relax
\typeout{** WARNING: IEEEtran.bst: No hyphenation pattern has been}%
\typeout{** loaded for the language `#1'. Using the pattern for}%
\typeout{** the default language instead.}%
\else
\language=\csname l@#1\endcsname
\fi
#2}}
\providecommand{\BIBdecl}{\relax}
\BIBdecl

\bibitem{5GEditorial14}
J.~Thompson, X.~Ge, H.-C. Wu, R.~Irmer, H.~Jiang, G.~Fettweis, and S.~Alamouti,
  ``5{G} wireless communication systems: prospects and challenges [{G}uest
  {E}ditorial],'' \emph{{IEEE} Communications Magazine}, vol.~52, no.~2, pp.
  62--64, Feb. 2014.

\bibitem{OMAvsNOMA}
P.~Wang, J.~Xiao, and L.~Ping, ``Comparison of orthogonal and non-orthogonal
  approaches to future wireless cellular systems,'' \emph{{IEEE} Vehicular
  Technology Magazine}, vol.~1, no.~3, pp. 4--11, Sept. 2006.

\bibitem{LTEA_Ghosh}
A.~Ghosh, R.~Ratasuk, B.~Mondal, N.~Mangalvedhe, and T.~Thomas,
  ``{LTE}-{A}dvanced: next-generation wireless broadband technology,''
  \emph{{IEEE} Wireless Communications Magazine}, vol.~17, no.~3, pp. 10--22,
  June 2010.

\bibitem{Trans_Por_Adap}
J.~Jang and K.~B. Lee, ``Transmit power adaptation for multiuser {OFDM}
  systems,'' \emph{{IEEE} Journal on Selected Areas in Communications},
  vol.~21, no.~2, pp. 171--178, Feb. 2003.

\bibitem{OFDMA_Optimality}
H.~Li and H.~Liu, ``An analysis of uplink {OFDMA} optimality,'' \emph{{IEEE}
  Transactions on Wireless Communications}, vol.~6, no.~8, pp. 2972--2983, Aug.
  2007.

\bibitem{Tse_Polymatroid}
D.~Tse and S.~Hanly, ``Multiaccess fading channels-{P}art {I}: Polymatroid
  structure, optimal resource allocation and throughput capacities,''
  \emph{{IEEE} Transactions on Information Theory}, vol.~44, no.~7, pp.
  2796--2815, Nov. 1998.

\bibitem{WeiYu_IWFJour}
W.~Yu, W.~Rhee, S.~Boyd, and J.~Cioffi, ``Iterative water-filling for gaussian
  vector multiple-access channels,'' \emph{{IEEE} Transactions on Information
  Theory}, vol.~50, no.~1, pp. 145--152, Jan. 2004.

\bibitem{IDMA_LiPing}
L.~Ping, L.~Liu, K.~Wu, and W.~K. Leung, ``Interleave division
  multiple-access,'' \emph{{IEEE} Transactions on Wireless Communications},
  vol.~5, no.~4, pp. 938--947, April 2006.

\bibitem{LDS_Editorial}
M.~Al-Imari, M.~Imran, and R.~Tafazolli, ``Low density spreading multiple
  access,'' \emph{Journal of Information Technology and Software Engineering},
  vol.~2, no.~4, pp. 1--2, Sep. 2012.

\bibitem{CodSpread_tradeoff}
V.~V. Veeravalli and A.~Mantravadi, ``The coding-spreading tradeoff in {CDMA}
  systems,'' \emph{{IEEE} Journal on Selected Areas in Communications},
  vol.~20, no.~2, pp. 396--408, Feb. 2002.

\bibitem{IMT_Guideline}
ITU, \emph{Guidelines for Evaluation of Radio Transmission Technologies for
  {IMT}-2000}.\hskip 1em plus 0.5em minus 0.4em\relax Recommendation ITU-R
  M.1225, 1997.

\bibitem{EURASIP_OFDMA}
M.~Al-Imari, P.~Xiao, M.~Imran, and R.~Tafazolli, ``Low complexity subcarrier
  and power allocation algorithm for uplink {OFDMA} systems,'' \emph{EURASIP
  Journal on Wireless Communications and Networking}, vol.~98, no.~1, pp. 1--6,
  April 2013.

\bibitem{TR25_996}
3GPP TR 25.996, ``Spatial channel model for Multiple Input Multiple Output
  (MIMO) simulations'', Release 11, Sept. 2012.

\bibitem{Jain_Fairness}
R.~K. Jain, D.-M.~W. Chiu, and W.~R. Hawe, ``A quantitative measure of fairness
  and discrimination for resource allocation in shared computer systems,''
  \emph{DEC Technical Report 301}, Sept. 1984.

\end{thebibliography}
\end{document}